\documentclass[aps,10pt,twocolumn,pra,citeautoscript,amsmath,amssymb,floatfix,nofootinbib,superscriptaddress,longbibliography]{revtex4-2}
\usepackage[left=2cm,right=2cm,top=2cm,bottom=2cm]{geometry}

\usepackage[utf8]{inputenc}
\usepackage[english]{babel}
\usepackage[T1]{fontenc}
\usepackage{siunitx}

\usepackage{braket}
\usepackage{ragged2e}
\usepackage[normalem]{ulem}
\usepackage{graphicx}
\usepackage[caption=false]{subfig}

\usepackage{amsmath}
\usepackage{amsfonts}
\usepackage{amssymb}

\usepackage{hyperref}
\usepackage{cleveref}
\usepackage{natbib}
\begin{document}

\title{Angular spectrum influence and entanglement characterization of Gaussian-path encoded photonic qudits}

\author{Borges, G. F.}
\email{Corresponding author:gfbj@fisica.ufmg.br}
\affiliation{Departamento de F\'{\i}sica, Universidade Federal de Minas Gerais,
 Caixa Postal 702, Belo Horizonte, MG 30123-920, Brazil}
\affiliation{Instituto de F\'isica, Universidade Federal de Uberl\^andia, 38400-902, Uberl\^andia, MG, Brazil
}
\author{Baldij\~ao, R. D.}
\affiliation{Departamento de F\'{\i}sica, Universidade Federal de Minas Gerais,
 Caixa Postal 702, Belo Horizonte, MG 30123-920, Brazil}
\affiliation{Instituto de F\'isica Gleb Wataghin, Universidade Estadual de Campinas, Campinas, SP 13083-859, Brazil}
\author{Matoso, A. A. }
\affiliation{Departamento de F\'{\i}sica, Universidade Federal de Minas Gerais,
 Caixa Postal 702, Belo Horizonte, MG 30123-920, Brazil}
\author{P\'adua, S.}
\affiliation{Departamento de F\'{\i}sica, Universidade Federal de Minas Gerais,
 Caixa Postal 702, Belo Horizonte, MG 30123-920, Brazil}
 




\begin{abstract}
Entangled quantum states play an important role in quantum information science and also in quantum mechanics fundamental investigations. Implementation and characterization of techniques allowing for easy preparation of entangled states are important steps in such fields. Here we generated entangled quantum states encoded in photons transversal paths, obtained by pumping a non-linear crystal with multiple transversal Gaussian beams. Such approach allows us to generate entangled states of two qubits and two qutrits encoded in Gaussian transversal path of twin photons. We make a theoretical analyses of this source, considering the influence of the pump angular spectrum on the generated states, further characterizing those by their purity and entanglement degree.
Our experimental results reveals that the generated states presents both high purity and entanglement, and the theoretical analysis elucidates how the pump beams profile can be used to manipulate such photonic states. 
\end{abstract}

\maketitle



\section{Introduction}

Encoding quantum states in real physical systems is a mandatory step to any implementation which aims to explore the quantum realm. Therefore, new ways to generate and encode quantum systems may serve to construct novel experiments and to find new paths towards feasible quantum machines. In particular, to be capable of easily generating entangled states is important, since this non-classical feature is responsible for many of the quantum advantages in quantum information processing \cite{erhard2020advances}. Here, we analyse in detail a source of photonic qudits, encoded in the transverse path degree of freedom, designed to produce entangled states of twin photons produced in the process of spontaneous parametric down conversion (SPDC).

The encoding of discrete path states in photons generated by SPDC is not a new idea; for instance, it was done by aligning fibers or slits  at {the photon paths} after the crystal \cite{walborn2010spatial,borges2014quantum,Peeters1,carvalho1,Neves1,Neves2,art:GlimaPRA}, as well as coupling silicon chips or photonic circuits in glass \cite{Wang2020_SchiarrinoChip} with defined path states. These strategies, however, suffer with some obstacles: either difficulties on efficiently and freely implementing operations on such systems (as it is the case for fiber or chip-based strategy) or difficulty on propagating the states through interferometers (the case of the slit strategy). Another crucial step in these sources is to transfer the entanglement of the twin photons to the discrete encoding, while preserving coherence with low loss. In the source presented here, we aim to encompass all such difficulties.

The main idea is to impose the discretization on the pump beam, before the crystal used for generating photon pairs by SPDC. If the right conditions are fulfilled, the twin photons will naturally carry the pump' s discretization while  entangled. Since manipulating the pump coherently is much easier than doing so with the twin photons, this also facilitates to execute different state preparations. By not using apertures to define the path-basis states one can also ensure lesser loss -- and thus increase efficiency of state generation.
In our implementation, the pump is initially in a Gaussian mode and is then split into several paths in one transverse direction, which will encode the path-basis states.  These paths will hit the same crystal in order to generate the twin photons entangled in path states defined by the pump directions. 

Similar strategies have been implemented recently and proven useful \cite{art:Guo2020PRL,art:hu2016contextpathst,art:CorrelationPath2018}. However, a detailed analysis of this kind of source, allowing for an understanding on what is the role of the pump's angular spectrum was still lacking. For instance, how does the profile of the pump affect the generated state?

I order to assure versatility for future endeavors -- not only to specific experiments -- the main features of whatever source is used to generate the states must be well known. For this purpose, here we make a characterization of this kind of sources, analysing the transfer of angular spectrum of the pump to the two-photons state. Such theoretical description makes explicit which pump beam parameters can be manipulated in order to control the two-photons states, thus providing a useful toolbox to perform engineering of this type of photonic quantum states.

Our theoretical analysis is focused on a Gaussian spatial mode, but can be easily adapted for arbitrary eigenstates of the propagation operator in the paraxial approximation, thus facilitating to reach higher-dimensional states and/or to create hyper-entangled states. By using only one transverse dimension to encode the path basis, we allow for the use of the techniques described in \cite{art:expAutomatedOper,art:ThoreticalAutomatedOper} to perform automated transformations in path encoded qudits. Finally, we describe in detail an experimental implementation of such a source, showing the presence of entanglement for both qubits and qutrits via measurements on both image and interference planes.
\section{Theoretical background}
\label{theory}

\begin{figure}[htbp]
\begin{center}
\fbox{\includegraphics[width=\linewidth]{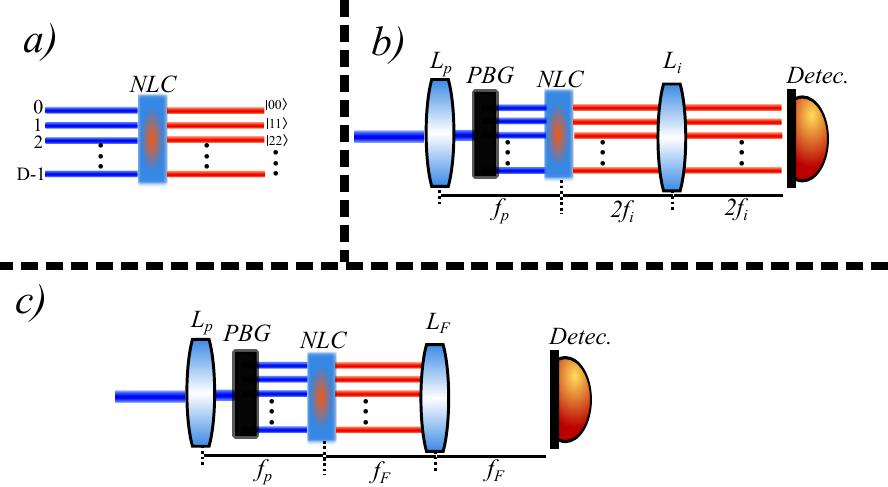}}
\caption{A schematic view of a setup able to generate two qudits in transverse photon path degrees of freedom. a) $D$ parallel laser beams pump a non linear crystal (NLC) giving rise to parametric down converted photon pairs in $D$ possible transverse Gaussian spatial modes. b) A spherical lens ($\rm L_{p}$) is placed before the parallel beams generator (PBG) with the NLC at the focal distance of the lens. Another spherical lens ($\rm L_{i}$) is placed between the crystal and the detection planes, at equal distances of twice the focal length. This optical configuration projects an unmagnified image of the crystal plane at the detection one. c) The spherical lens ($\rm L_{F}$) performs the optical Fourier transform of the crystal plane at the detection plane. Such configuration allows the measurement of the two-photons path interference patterns.}
\label{fig:schematic}
\end{center}
\end{figure}

The idea behind this approach is sketched in Fig.\ref{fig:schematic}(a), where a type II non linear crystal is coherently pumped by $D$ idealized  laser beams in the monochromatic approximation with frequency $\omega_{p}$ \cite{walborn2010spatial}. The pump beams intensities of the $D$ spatial modes are sufficient to generate just a single photon pair by SPDC. In other words,  the probabilities of two pairs conversion by the same beam or the simultaneous pair conversion by different beams are negligible. In addition the phase-matching of the non-linear crystal is designed for collinear generation of equal frequency photons, with frequency $\omega = \omega_{p} /2$. The photon pairs are selected by using narrow interference filters after the crystal (not shown in Fig.\ref{fig:schematic}(a)). In this way, such approach is able to create a coherent superposition of $D$ possible photon paths in which there is a high photon pair generation in the $D$ Gaussian modes. The resulting quantum state in the transverse path degree of freedom, regardless a vacuum state that contributes with no photon pair coincidence detection, is given by:

\begin{equation}
\label{eq:statepictor}
\ket{\Psi} = \sum_{\ell=0}^{D-1} \alpha_{\ell} \ket{\ell,\ell}\:,
\end{equation} 

\noindent
where $\sum|\alpha_{\ell}|^2=1$ and $\ket{\ell}$ is a path state labelled by $\ell=0,1,...,D-1$, as shown in Fig.\ref{fig:schematic}(a). In the next subsection we show how this setup allows for the  construction of states represented by \eqref{eq:statepictor}, which might be a maximally entangled state in the transverse paths if $|\alpha_{\ell}|=1/\sqrt{D}$ for all $\ell$.

In order to ensure the superposition shown in \eqref{eq:statepictor}, we need to generate spatially and temporally coherent $D$ parallel pump beams. These parallel pump beams can be obtained by using beam displacers (BD), half wave plates (HWP), and quarter wave plates (QWP). A BD is a birefringent linear optical device that acts as a Polarizing Beam Splitter where the output beams follow parallel paths and have orthogonal polarization. The exiting beams are commonly referred as ordinary ray (\textit{o}-ray) and extraordinary ray (\textit{e}-ray), depending on how each beam's polarization is oriented with respect to the crystal optical axis \cite{liv:hecht2002optics}.
In the experimental setup reported here the \textit{o}-ray is horizontally polarized while the \textit{e}-ray is vertically polarized. By using a HWP on the incident beam we are able  to control the relative intensity of the parallel beams. The coherence condition is guaranteed, because the wave packet temporal coherence of continuous wave (CW) lasers are usually much larger than the longitudinal walk-off caused by BDs and, therefore,  it is possible to obtain parallel coherent pump beams. This use of BDs and HWPs -- or any other device capable of producing coherent parallel beams, e.g \cite{art:torres} --  is referred here as a \textit{Parallel Beam Generator} (PBG) and the separation between paths will be denoted by $d$.

Up to now, each pump and photon beam paths are considered as a ray, disregarding the transverse field profile. 
A more realistic theoretical description to this path-entangled states needs to consider the profile of the pump beams, which can be taken as Gaussian in the monochromatic paraxial approximation. This is essential if one needs to consider propagation or interference of the different path components, as, for example, in the complete state characterization.

Indeed, the pump transverse profile affects the generated state, as well as the photon paths correlations observed at the crystal and Fourier planes (this is an usual way to verify non-classical correlations of spatial qudit states \citep{borges2014quantum,Neves1}). 
These aspects will be discussed in the following subsections.

\subsection{The pump angular spectrum influence} \label{subsec:2a}

In order to ensure distinguishability between the path modes it is necessary to avoid transverse overlap between the parallel pump beams. This imposes a restriction on how close the beams can be, given their width. However, they cannot be excessively separated, since they must fit into the crystal transverse size. Both distinguishability and compactness can be achieved by decreasing the pump Gaussian width at the crystal plane. The adopted strategy was to use a spherical lens with focal length $f_{p}$ before the PBG so that the crystal is placed at the focal region. This situation is shown in Fig.\ref{fig:schematic}(b). Restricting our analysis to one transverse direction ($x$-direction), the spatial field distribution profile of the pump beam can be expressed by:

\begin{equation}
\label{eq:gaussprofile}
E\left( x,z \right) = \sum_{\ell=0}^{\left(D-1\right)} E_{0}^{\left(\ell\right)} e^{-\frac{\left(x-\ell d\right)^2}{w^2\left(z\right)}}
e^{\left[-i\left(kz+\frac{k(x-\ell d)^2}{2R^{\prime}(z)}-\zeta(z)\right)\right]} \:,
\end{equation}

\noindent
where $k$ is the pump wavenumber, $E_{0}^{\left(\ell\right)}$ is the amplitude of the electric field on path $\ell$, $w\left(z\right)$ is the beam width, $R^{\prime}(z)$ is the radius of curvature transmitted by the spherical lens, which is related to the incident beam curvature $R$ and the lens focal length $f_{P}$ through the expression $1/R^{\prime}=1/R-1/f_{p}$ and $\zeta(z)$ is the Gouy phase \cite{teich1991fundamentals}. Here we suppose that $R(z)$, $w(z)$ and $\zeta(z)$ are equal for all values of $\ell$.

The beam waist after the lens is given by the parameter $w^{\prime}_0=w_{p}/ \sqrt{1+\left(\pi w_{p}^{2} /\lambda R^{\prime}\right)^2}$, being $w_p$ the width of the incident beam on the lens $L_p$.

Since the pump angular spectrum is transferred to the down converted photons state \cite{PhysRevA.57.3123}, the desired two-photons quantum state in the transverse momentum representation is given by

\begin{equation}\label{eq:generatedstate}
\begin{split}
\ket{\Psi_q} \propto& \sum_{\ell=0}^{D-1}A_{\ell} \iint\limits_{-\infty}^{\infty} dq_{i} dq_{s} \exp{\left[\frac{-w^{\prime 2}(z) \left(q_{s}+q_{i})^2\right)}{4}\right]} \times \\ 
 &\times e^{-i \ell d\left(q_{i}+q_{s}\right)}\ket{1q_{s},H}\otimes\ket{1q_{i},V} \:.
\end{split}
\end{equation}

\noindent
where $A_{\ell}$ are the amplitude probabilities of detecting both photons at the path $\ell$, H and V are respectively the horizontal and vertical polarization components, $1/w^{\prime 2}(z)=1/w^2(z)+ik/2R^{\prime}(z)$ and $q_i$ ($q_s$) is the transverse momentum component of the idler (signal) photon. For a well collimated incident beam on the lens $L_p$, it is reasonable to assume that the focusing plane is equal to the plane where the beam waist $w'_0$ is located. Once we place the nonlinear crystal in this plane where $w^{\prime}(z) = w_0^{\prime}$, the generated photon pair state becomes

\begin{equation}\label{eq:generatedstate2}
\begin{split}
\ket{\Psi_q} \propto& \sum_{\ell=0}^{D-1}A_{\ell} \iint\limits_{-\infty}^{\infty} dq_{i} dq_{s} \exp{\left[\frac{-w_{0}^{\prime 2} \left(q_{s}+q_{i})^2\right)}{4}\right]} \times \\ 
 &\times e^{-i \ell d\left(q_{i}+q_{s}\right)}\ket{1q_{s},H}\otimes\ket{1q_{i}, V} \:.
\end{split}
\end{equation}

By using the transverse momentum-position Fock state transform, in which $\ket{1q}=\int dx \: e^{-iqx} \:\ket{1x}$, we are able to write the two-photons state in the transverse position representation \cite{art:Saleh2001}:

\begin{equation}\label{eq:positgeneratedstate}
\ket{\psi_x} \propto \sum_{\ell=0}^{D-1}A_{\ell} \int\limits_{-\infty}^{\infty} dx \exp\left[\frac{-\left(x-\ell d\right)^2}{2w^{\prime 2}_{0}}\right]\ket{1x,H}\otimes\ket{1x,V} \:.
\end{equation}

The above expression can be viewed as a generalization of \eqref{eq:statepictor}, in which the spatial two-photons mode $\ket{\ell,\ell}$ is now represented in the photon transverse position representation. The orthogonality between two distinct states that form the discrete path state base occurs when the spatial overlapping of the Gaussian modes for different $\ell$s is negligible. We can achieve this by choosing $f_{p}$ such that the beam width is smaller than the adjacent mode separation $d$. 

The spatial profile of the generated photon pairs follow the same propagation properties of the pump Gaussian beams\cite{walborn2010spatial,PhysRevA.57.3123}. Since the pump beam is focused at the crystal plane, the different path modes of the down-converted photons diverge while they propagate in free space causing the overlapping of the down converted beams before the detection plane. To avoid this, it is necessary to use an extra spherical lens ($\mathrm{L_i}$) between the crystal plane and the detection plane, in a $2f-2f$ lens configuration to provide an unmagnified image of the photon pairs generation plane (Fig.\ref{fig:schematic}(b)). The coincidence count probabilities in the $2f-2f$ configuration are obtained by propagating the electric field operator from the crystal plane through a spherical lens to the detection plane and calculating the forth-order correlation in the fields (second-order correlation in the intensities).

The two-photons count probability is given by:
\begin{equation}\label{eq:imagecoincidenceprob}
    \mathcal{P}\left( x_{i}=x_{s}=x\right) \propto \left|e^{\frac{ik}{f}x^2}\sum_{\ell=0}^{D-1} A_{\ell} \exp\left[-\frac{\left(x-\ell d \right)^2}{w_{0}^{\prime 2}} \right] \right|^{2}.
\end{equation} 

It is important to point out that the use of a spherical lens at the pump beam is unnecessary if the pump beam separation provided by the PBG is large enough to avoid the overlaping of the pump beams at the crystal plane. 
 
\subsection{Conditional interference pattern}

Previous works that prepared and characterized two-photons spatial qudit states measured the conditional  transverse interference pattern aiming to relate such conditionality with the quantum character of the spatial correlations in these systems \cite{borges2014quantum,carvalho1}. Furthermore some of them used the two-photons interference pattern to realize fundamental tests on quantum mechanics and characterize these quantum systems \cite{PhysRevA.60.1530,Pierre1,carvalho1,Peeters1,art:Sinha418}. 

An useful approach to obtain the conditional interference pattern is to project the Optical Fourier transform of the crystal plane at the detection plane (Fourier plane). Experimentally this can be done by using a spherical lens after the crystal in a $f-f$ configuration, as shown in the scheme presented in Fig.\ref{fig:schematic}(c). 

By propagating the electric field operators from the crystal plane to the Fourier plane,  and by supposing that the idler and the signal photons are separated spatially by a polarizer beam splitter (Fig.\ref{fig:experimental1}(b)) one can find the coincidence count probability at this plane \cite{Neves1,liv:goodman2005}: 

\begin{equation}\label{eq:conditionalpatt}
\begin{split}
\mathcal{P}\left( x_{i},x_{s}\right) \propto& \left|\sum_{\ell=0}^{D-1} A_{\ell}\left(\frac{2k}{f_F}\right)^2\exp{\left[-\frac{kw^{\prime 2}_{0}\left(x_{s}+x_{i}\right)^2}{f_F}\right]} \times \right. \\
&\left. \times \exp{\left[\frac{2ikd}{f_F}\ell   \left(x_{s}+x_{i}\right)\right]} \right|^2,
\end{split}
\end{equation}

\noindent
in the monochromatic, thin crystal and paraxial approximations. \eqref{eq:conditionalpatt} shows an interference pattern that depends on the sum of the detectors position ($x_i$ and $x_s$), which is the reason why such pattern is called conditional.
In addition \eqref{eq:conditionalpatt} elucidates the role of the pump beam waist in the photon pair interference pattern. 
The waist of the pump Gaussian beams affect the diffraction envelope of the conditional two-photons interference pattern.

\section{Experimental implementation and characterization}
 
\begin{figure}[htbp]
\centering
\fbox{\includegraphics[width=\linewidth]{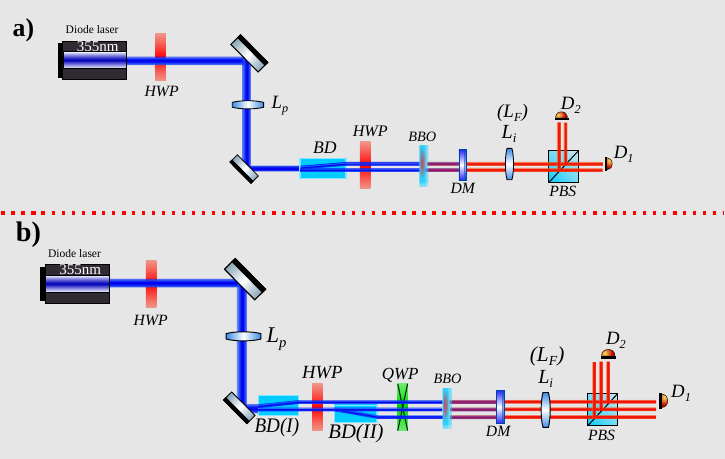}}
\caption{Experimental configuration used to generate entangled a)two-qubit and b)two-qutrit states in the transverse path degree of freedom. In both configurations the nonlinear type II BBO crystal was set for collinear SPDC generation and for horizontally polarized pump beam. The vertical polarization component of each pump beam does not generate any photon pairs due to the phase-matching condition.}
\label{fig:experimental1}
\end{figure}

Fig.\ref{fig:experimental1} shows a possible experimental setup which can implement the spatial qudits source  discussed above. In this setup a $\SI{355}{\nano\meter}$ CW laser provides a Gaussian beam with $\SI{90}{\milli \watt}$. The linear polarization of the laser beam is controlled by a half-wave plate, and this beam emerges from it with diagonal polarization at $\SI{+45}{\degree}$ for the (a) case, resulting in equal intensities for the two orthogonal polarization components emerging from the first calcite beam displacer. Our PBG is composed by a sequence HWP-BD-HWP for the two-qubit case and HWP-BD-HWP-BD-QWP for the two-qutrit case, as shown in Fig.\ref{fig:experimental1}. The transversal displacement provided by our BDs is $d\approx\SI{1}{\milli \meter}$ and the orientation of the waveplates determines the modulus of each $A_{\ell}$.

A type II BBO, $\SI{5}{\milli \meter}$ long crystal  was setted for horizontal polarization phase matching, \textit{i.e}, only the horizontal component of the pump beams are able to generated down converted photon pairs with orthogonal polarization. Another important characteristic of the  down converted photons is that they are collinear with the pump beams, and the central generated wavelength is $\SI{710}{\nano\meter}$. As mentioned in section \ref{theory}, four photon generation processes -- two pairs conversion by the same beam or the simultaneous one pair conversion by different beams -- are negligible in this setup, since we are using low intensity CW laser as a pump beam.

In order to reduce the Gaussian widths at the crystal plane we used a spherical lens with focal length $f_{p} = \SI{50}{\centi\meter}$. The crystal is placed at the focal distance from the lens. The large focal length ensures incident beams on the crystal with low divergence. The spatial correlation between the generated photons follows the pump beam divergence profile due to the angular spectrum transference from the pump beam to the generated photon pairs \cite{PhysRevA.57.3123}. A dichroic mirror after the crystal reflects back the pump beams and transmits the down converted photons.

In this experiment, we used a lens with focal length $f_i = \SI{25}{\centi\meter}$ in a $2f-2f$ configuration to obtain an image of the crystal plane at the detection plane. As discussed in subsection \ref{subsec:2a}, this optical configuration propagates the two-photons state from the crystal plane (\eqref{eq:generatedstate}) to the detection plane maintaining its original form \cite{art:GlimaPRA}. This allows access to correlations in the computational basis of the implemented state.

After the lens $L_i$ the orthogonally polarized photons are split by a polarizing beamsplitter (PBS) and directed to two avalanche photodiode detectors (APD), at transverse positions denoted by $x_1$ and $x_2$, which are free to vary. A $\SI{100}{\micro\meter}$ single slit and an interference filter centered at $\SI{710}{\nano \meter}$ with $\SI{10}{\nano \meter}$ bandwidth are placed in front of each detector to spatially select the frequency degenerated photon pairs. The single and coincidence counts are acquired and recorded by a coincidence circuit with a $\SI{5.4}{\nano\second}$ temporal window.

\begin{figure}[h!]
 \includegraphics[width=\linewidth]{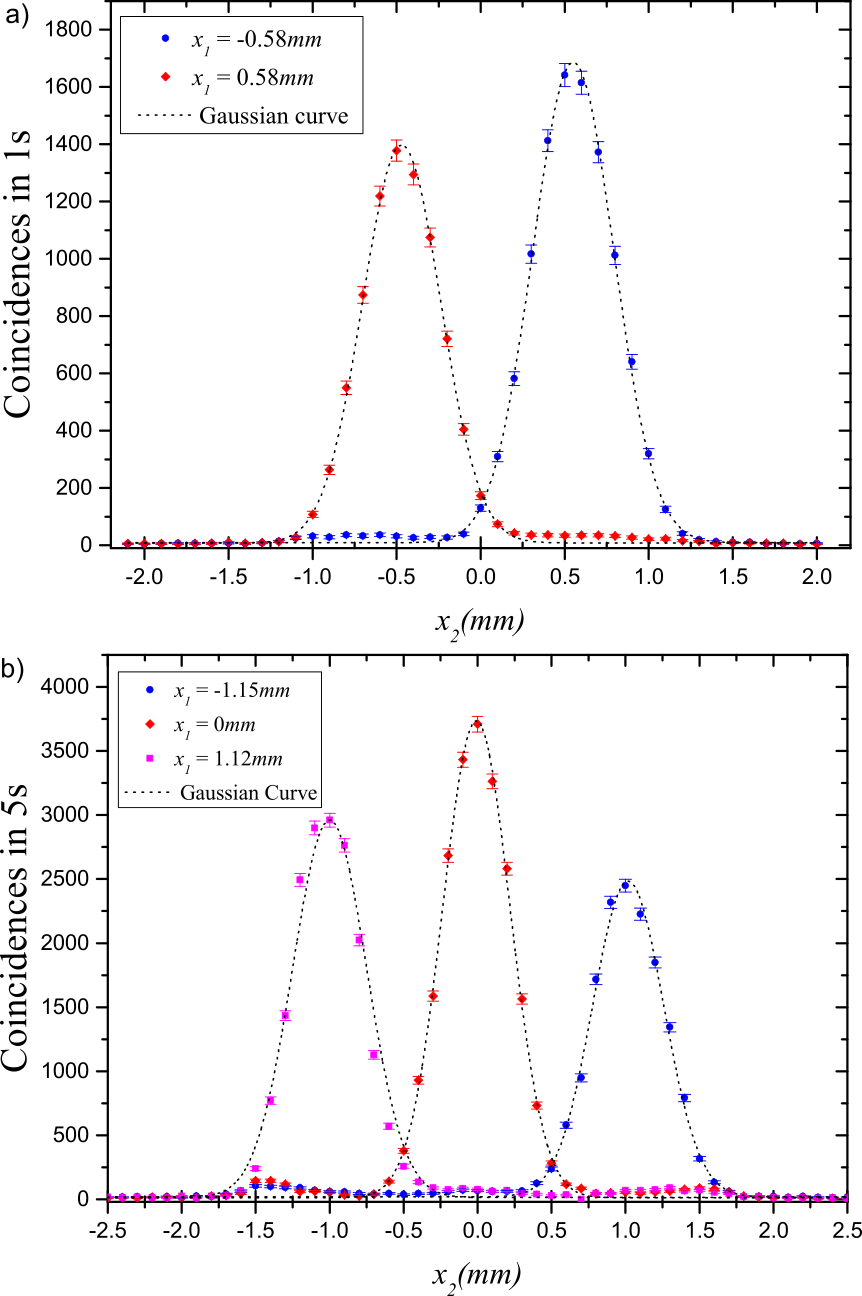}
    \caption{Coincidence count distribution at the image plane ($2f-2f$ configuration) for the a) two-qubit and b) two-qutrit state. The position of detection $1$ is kept fixed in different values while the position of detector $2$ is varied in the x-direction in constant steps. Each set of experimental points -- blue filled circles and red rhombus in a) or blue filled circles, red rhombus and magenta squares in b) -- is measured for a given fixed position of detector $1$.}
    \label{fig:imagem}
\end{figure}

By maintaining the position of detector $1$ fixed ({$x_1$ constant}), and varying the position of detector $2$ we obtain the coincidence count distribution at the image plane that corresponds to photon pair distribution at the crystal plane. Such distribution is proportional to  \eqref{eq:imagecoincidenceprob}, and reaches the coincidence counts peak when $x_2\approx x_1$, as shown in Fig.\ref{fig:imagem} . However, due to Gaussian path overlaps a small amount of coincidence counts are observed when the  detectors 1 and 2 are placed in positions related to Gaussian paths with different $\ell$s. This overlap implies that there exists a non-null probability $\left|\alpha_{ij}\right|^2$ of detecting one photon in a path indexed by $i$  while the another one (from the same pair) is in the path indexed by $j$, with $i(j)=0,1$ for two-qubit. In this way, the Gaussian paths overlap causes an experimental deviation from the ideal state previewed in \eqref{eq:statepictor}.

\begin{figure}[h!]
    \centering
 \includegraphics[width=\linewidth]{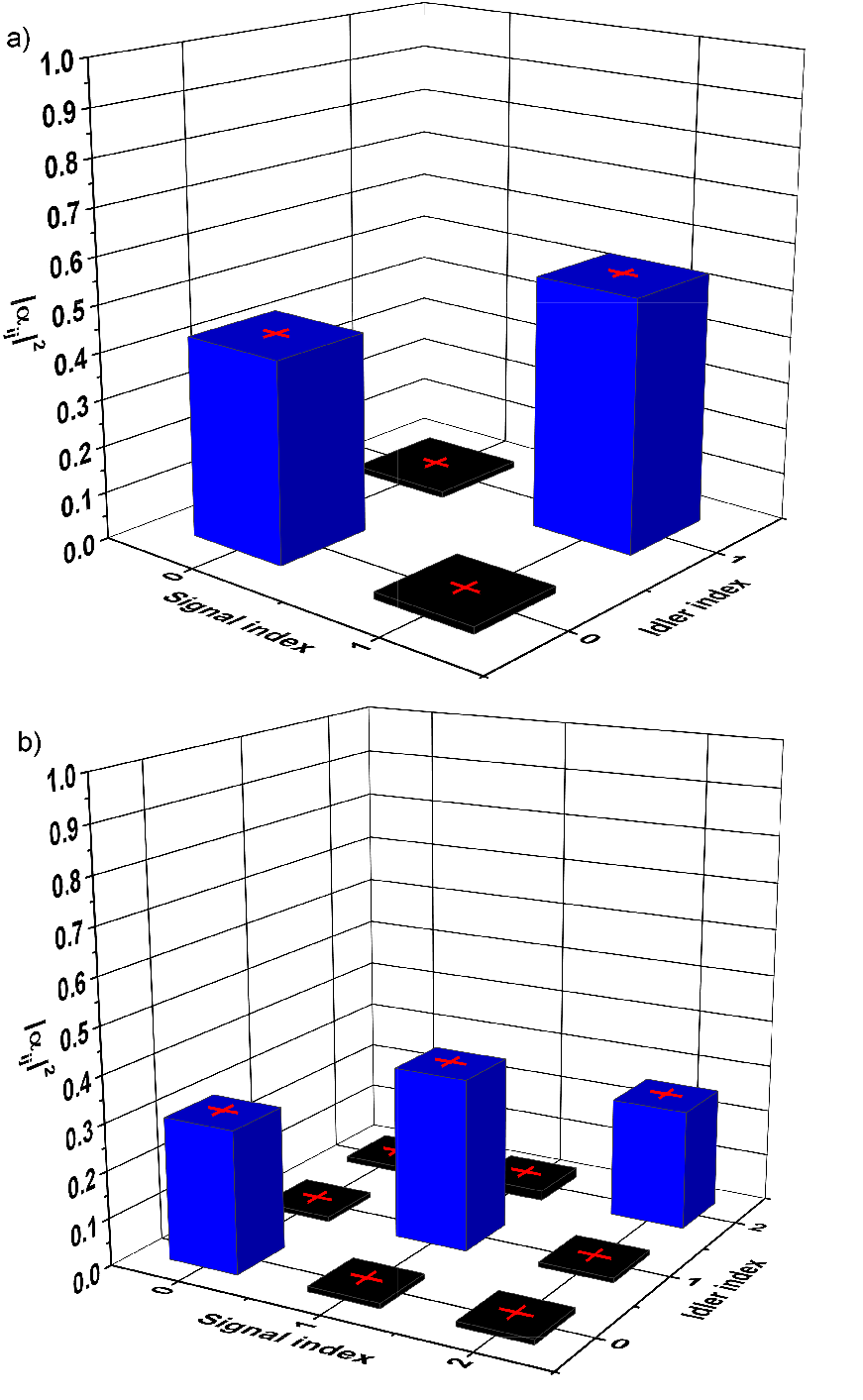}
 \caption{Probabilities $\left|\alpha_{ij}\right|^2$ of detecting one photon from the pair in the path $i$ and the another one in the path $j$, for a) two-qubit (where $i,j=0,1$) and b) two-qutrit (where $i,j=0,1,2$). Such probabilities are obtained by propagating the photons through a $2f-2f$ lens configuration and detecting the pair at the image plane. 
}
\label{fig:imagemCoeff}
\end{figure}

In order to determine the probabilities $\left|\alpha_{ij}\right|^2$ firstly is necessary to identify each Gaussian peak of the coincidence counts spatial distribution with a $(ij)$ pair and evaluate the area under each one. The probabilities are given by the normalized area of each Gaussian peak $(ij)$. Here we normalize the Gaussian peak areas by the total area,\textit{i.e}, the summation of all $(ij)$ Gaussian areas). The resultant probabilities are shown in Fig.\ref{fig:imagemCoeff}. Note that the probabilities $\left|\alpha_{ij}\right|^2$ for $i\neq j$ are negligible compared with the probabilities for $i=j$, so one can conclude that the overlapping between the Gaussian paths is also negligible. Such conditional behavior of the coincidence counts allows one to infer that the relevant path correlations are the same described by the state in \eqref{eq:statepictor}, so we can take $\alpha_{ij} \equiv \alpha_{\ell}$ for $i=j$ and $\alpha_{ij} \equiv 0$, for $i\neq j$.

On the other hand the quantumness of such correlation can be assured just by observing the conditional interference patttern \cite{borges2014quantum,Neves1}.

\begin{figure}[h!]
    \centering
 \includegraphics[width=\linewidth]{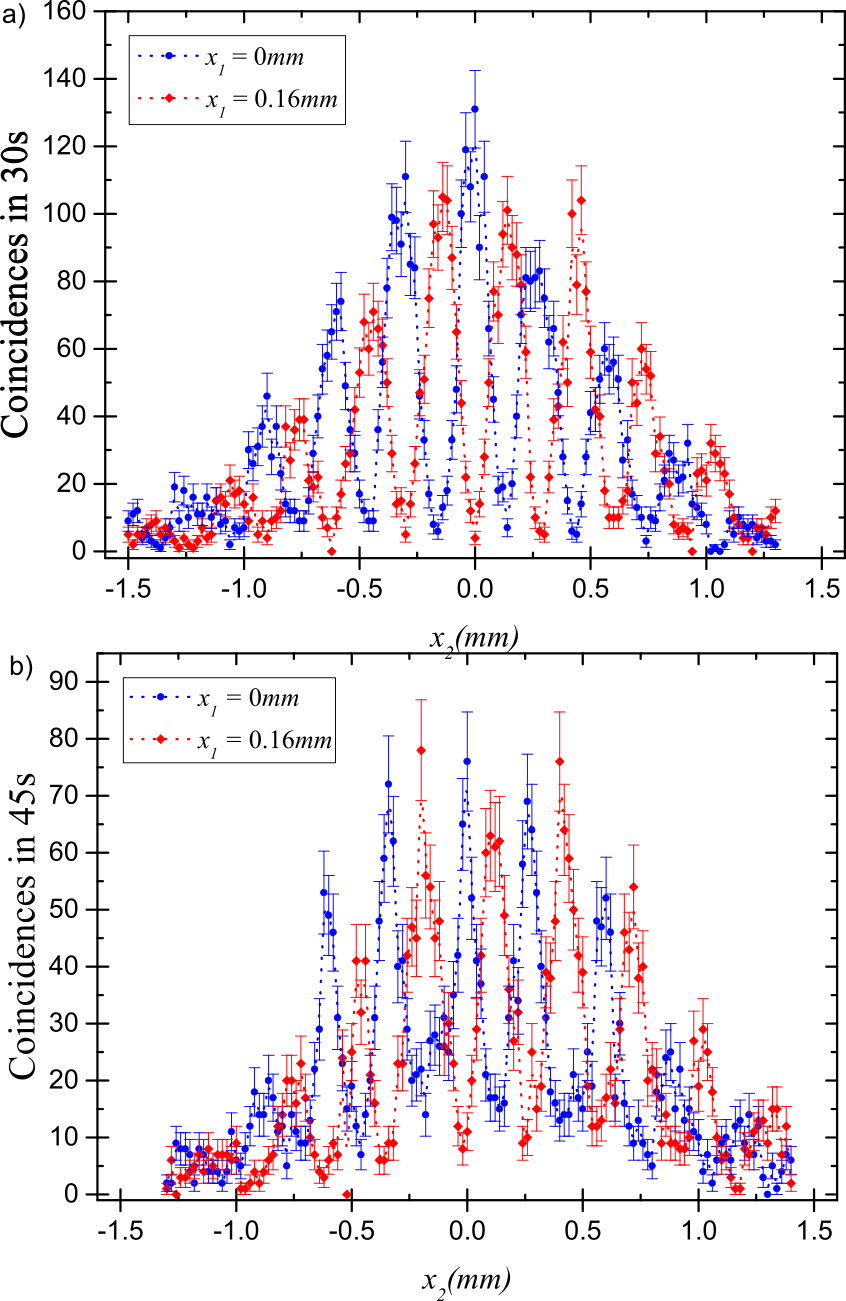}
    \caption{Conditional interference pattern, for a) two-qubit and b) two-qutrit, obtained by propagating the photons through a $f-f$ lens configuration and detecting the pair at the Fourier Plane with detector 1 at a fixed position. One can note that the fixed position of detector $1$ acts as a phase for the two-photons interference pattern as predicted by \eqref{eq:conditionalpatt}.}
    \label{fig:interference}
\end{figure}

The conditional interference pattern (CIP) was obtained by projecting the optical Fourier transform of the crystal plane at the detection plane by using a lens of focal length $f_{F} = \SI{50}{\centi\meter}$ in the $f-f$ configuration. The detector's slits were changed from $\SI{100}{\micro \meter}$  to $\SI{50}{\micro \meter}$. Fig.\ref{fig:interference}(a) shows two CIPs, for the two-qubit configuration. The coincidence counts are acquired first by setting the position of one detector while the other detector scans the pattern transversely with steps of $\SI{0.20}{\milli \meter}$. This procedure is repeated for different positions of the first detector.

In a similar way we can observe the generation of two entangled qutrits with the experimental scheme show in Fig.\ref{fig:experimental1}(b). Now, an additional BD is necessary to generate three pump beams, as follows. First, the HWP(I) is set to transform the laser polarization in a superposition of horizontal and vertical components, such that the relative intensity is $\frac{2}{3} : \frac{1}{3}$, respectively. After the first BD, we have two beams with different intensities. Then, these two unbalanced laser beams pass through the HWP(II), which is oriented at $\SI{22.5}{\degree}$ with respect to the horizontal. Thus, each beam polarization becomes $\SI{-45}{\degree}$ (more intense beam) and $\SI{+45}{\degree}$ (less intense beam). Finally, the $\SI{-45}{\degree}$ linearly polarized beam pass through the BD(II), splitting into two equally-intense beams with horizontal and vertical polarization, while the other one does not cross this element. The three pump beams have now the same total intensity -- this is not enough yet, since only the horizontal component is able to generate the twin-photons.

A quarter wave plate (QWP) is inserted after BD(II), oriented at $\SI{45}{\degree}$ with respect to horizontal, aiming to have equal intensities in the horizontal polarization component of each pump beam. This transforms the horizontal and vertical polarized beams transmitted by BD(II) to right and left hand circular polarized beams, respectively. The polarization of the third beam not transmitted through BD(II) is not affected by the QWP. At this point our three pump beams have the same amount of horizontal polarization component leading to the same probabilities for the down conversion on each path.

The probabilities $\left|\alpha_{ij}\right|^2$ for the  two-qutrit path state are shown in Fig.\ref{fig:imagemCoeff}(b), with $i(j)=0,1,2$. The coincidence measurements used to calculate these probabilities were obtained in the same experimental conditions as the two-qubit case. One notes that the relevant probabilities occur when detectors register photons from the same path, $i=j$. In this way, following the same reasoning for the two-qubit case one can conclude that the two-qutrit path state has the same correlations shown in \eqref{eq:statepictor}, with $\alpha_{ij} \equiv \alpha_{\ell}$ for $i=j$ and $\alpha_{ij} \equiv 0$, for $i\neq j$.

The conditional interference pattern for the two-qutrit state was also obtained, using the same experimental procedure as for the two-qubit case. These two CIPs are shown in Fig.\ref{fig:interference}(b), where we can note some characteristic secondary maxima of the three Gaussian beams interference as predicted by \eqref{eq:conditionalpatt}.

We calculate the concurrence for two-qubit   \cite{art:SchimdtConcurrence,LeoNevesetalConcurrence,WootersConcurrence} and two-qutrit \cite{cereceda2003degree,art:Iconcurrenceintro} for quantifying the entanglement of the generated states.

As discussed above, the prepared states can be fairly described by \eqref{eq:statepictor},  making the Schmidt decomposition quite simple \cite{peres2006quantum,ekert1995entangled,LeoNevesetalConcurrence}. 
In other words, the Schmidt decomposition for this state is given by:
 \begin{equation}
 \label{eq:stateSchmidt}
 \ket{\Psi_{S}} = \sum_{\ell=0}^{D-1} \kappa_{\ell'} \ket{\ell',\ell'}\:,
\end{equation}
\noindent
with the Schmidt's coefficients given by $\kappa_{\ell'}=\left|\alpha_{\ell} \right|$ and the Schmidt basis $\{\ket{\ell',\ell'}\}$ differing from  $\{\ket{\ell,\ell}\}$ just by a phase $\phi_l$.

For two-qubit the concurrence is given by \cite{art:SchimdtConcurrence}:
\begin{equation} \label{eq:concQubits}
    C_{2 \times 2} = 2 \kappa_{0} \kappa_{1}\:.
\end{equation}

Using the square root of the probabilities $\left|\alpha_{00}\right|^2$ and $\left|\alpha_{11}\right|^2$ obtained above, the resulting concurrence is: $C_{2 \times 2} = 0.97 \pm 0.01$, which shows a high entanglement degree for the prepared two-qubit state.

For the two-qutrit state the concurrence can be calculated from the Schmidt's coefficients using the equation \cite{cereceda2003degree}:

\begin{equation} \label{eq:IconcQutrits}
C_{3 \times 3} = \sqrt{3\left(\kappa^{2}_{0}\kappa^{2}_{1} +\kappa^{2}_{1}\kappa^{2}_{2}+\kappa^{2}_{2}\kappa^{2}_{0}   \right)}\:,
\end{equation}
\noindent
with $\kappa_{\ell}$ obtained again from the probabilities $\left|\alpha_{\ell}\right|^2$. In this case the concurrence is: $C_{3 \times 3}=0.92 \pm 0.01$, also revealing a high degree of entanglement for the two-qutrit states.

\section{Conclusion}
This work provides a theoretical background and an experimental implementation of an entangled spatial photonic qudits source. We calculated the generated two-photons state by using the angular spectrum transference theory, which allowed us to also calculate the coincidence count probabilities for both the crystal and the Fourier plane. Entanglement is demonstrated in the photon paths degree of freedom defined by Gaussian modes. It was quantified calculating the concurrence for two-qubit and two-qutrit case, obtaining ($0.97 \pm 0.01$) and ($0.92 \pm 0.01$) respectively. The main advantage of this approach is the maintenance of the Gaussian beam profile throughout the propagation followed by the generated qudits. This provides technical facilities for coupling the spatial entangled qudits in optical fibers or in photonic circuits as well as for applying automated operations. Moreover, the theory and characterization developed here are easily extended to high-order Gaussian modes, which provides a perspective for reaching higher-dimensional entangled states or hyper-entangled states involving angular momenta. 

\section{Acknowledgements}
Authors thank the Brazilian agencies CNPq, FAPEMIG and CAPES for financial support.
This work is part of the Brazilian National Institute for Science and Technology in Quantum Information.
RDB acknowledges funding by S\~{a}o Paulo Research Foundation - FAPESP, grant no. 2016/24162-8.

\bibliographystyle{unsrtnat}
\bibliography{references}

\begin{thebibliography}{30}
\providecommand{\natexlab}[1]{#1}
\providecommand{\url}[1]{\texttt{#1}}
\expandafter\ifx\csname urlstyle\endcsname\relax
  \providecommand{\doi}[1]{doi: #1}\else
  \providecommand{\doi}{doi: \begingroup \urlstyle{rm}\Url}\fi

\bibitem[Erhard et~al.(2020)Erhard, Krenn, and Zeilinger]{erhard2020advances}
Manuel Erhard, Mario Krenn, and Anton Zeilinger.
\newblock Advances in high-dimensional quantum entanglement.
\newblock \emph{Nature Reviews Physics,}, pages 1--17, 2020.

\bibitem[Walborn et~al.(2010)Walborn, Monken, P{\'a}dua, and
  Ribeiro]{walborn2010spatial}
Stephen~P Walborn, CH~Monken, S~P{\'a}dua, and P~H~Souto Ribeiro.
\newblock Spatial correlations in parametric down-conversion.
\newblock \emph{Physics Reports}, 495\penalty0 (4):\penalty0 87--139, 2010.

\bibitem[Borges et~al.(2014)Borges, Carvalho, de~Assis, Ferraz, Ara{\'u}jo,
  Cabello, Cunha, and P{\'a}dua]{borges2014quantum}
Gilberto Borges, Marcos Carvalho, Pierre-Louis de~Assis, Jos{\'e} Ferraz,
  Mateus Ara{\'u}jo, Ad{\'a}n Cabello, Marcelo~Terra Cunha, and Sebasti{\~a}o
  P{\'a}dua.
\newblock Quantum contextuality in a young-type interference experiment.
\newblock \emph{Physical Review A}, 89\penalty0 (5):\penalty0 052106, 2014.

\bibitem[Peeters et~al.(2009)Peeters, Renema, and van Exter]{Peeters1}
W.~H. Peeters, J.~J. Renema, and M.~P. van Exter.
\newblock Engineering of two-photon spatial quantum correlations behind a
  double slit.
\newblock \emph{Phys. Rev. A}, 79:\penalty0 043817, Apr 2009.
\newblock \doi{10.1103/PhysRevA.79.043817}.
\newblock URL \url{http://link.aps.org/doi/10.1103/PhysRevA.79.043817}.

\bibitem[Carvalho et~al.(2012)Carvalho, Ferraz, Borges, de~Assis, P\'adua, and
  Walborn]{carvalho1}
M.~A.~D. Carvalho, J.~Ferraz, G.~F. Borges, P.-L de~Assis, S.~P\'adua, and
  S.~P. Walborn.
\newblock Experimental observation of quantum correlations in modular
  variables.
\newblock \emph{Phys. Rev. A}, 86:\penalty0 032332, Sep 2012.
\newblock \doi{10.1103/PhysRevA.86.032332}.
\newblock URL \url{http://link.aps.org/doi/10.1103/PhysRevA.86.032332}.

\bibitem[Neves et~al.(2005)Neves, Lima, Aguirre~G\'omez, Monken, Saavedra, and
  P\'adua]{Neves1}
Leonardo Neves, G.~Lima, J.~G. Aguirre~G\'omez, C.~H. Monken, C.~Saavedra, and
  S.~P\'adua.
\newblock Generation of entangled states of qudits using twin photons.
\newblock \emph{Phys. Rev. Lett.}, 94:\penalty0 100501, Mar 2005.
\newblock \doi{10.1103/PhysRevLett.94.100501}.
\newblock URL \url{http://link.aps.org/doi/10.1103/PhysRevLett.94.100501}.

\bibitem[Neves et~al.(2004)Neves, P\'adua, and Saavedra]{Neves2}
Leonardo Neves, S.~P\'adua, and Carlos Saavedra.
\newblock Controlled generation of maximally entangled qudits using twin
  photons.
\newblock \emph{Phys. Rev. A}, 69:\penalty0 042305, Apr 2004.
\newblock \doi{10.1103/PhysRevA.69.042305}.
\newblock URL \url{http://link.aps.org/doi/10.1103/PhysRevA.69.042305}.

\bibitem[Lima et~al.(2006)Lima, Neves, Santos, Aguirre~G\'omez, Saavedra, and
  P\'adua]{art:GlimaPRA}
G.~Lima, Leonardo Neves, Ivan~F. Santos, J.~G. Aguirre~G\'omez, C.~Saavedra,
  and S.~P\'adua.
\newblock Propagation of spatially entangled qudits through free space.
\newblock \emph{Phys. Rev. A}, 73:\penalty0 032340, Mar 2006.
\newblock \doi{10.1103/PhysRevA.73.032340}.
\newblock URL \url{https://link.aps.org/doi/10.1103/PhysRevA.73.032340}.

\bibitem[Wang et~al.(2020)Wang, Sciarrino, Laing, and
  Thompson]{Wang2020_SchiarrinoChip}
Jianwei Wang, Fabio Sciarrino, Anthony Laing, and Mark~G. Thompson.
\newblock Integrated photonic quantum technologies.
\newblock \emph{Nature Photonics}, 14\penalty0 (5):\penalty0 273--284, May
  2020.
\newblock ISSN 1749-4893.
\newblock \doi{10.1038/s41566-019-0532-1}.
\newblock URL \url{https://doi.org/10.1038/s41566-019-0532-1}.

\bibitem[Hu et~al.(2020)Hu, Xing, Liu, Huang, Li, Guo, Erker, and
  Huber]{art:Guo2020PRL}
Xiao-Min Hu, Wen-Bo Xing, Bi-Heng Liu, Yun-Feng Huang, Chuan-Feng Li, Guang-Can
  Guo, Paul Erker, and Marcus Huber.
\newblock Efficient generation of high-dimensional entanglement through
  multipath down-conversion.
\newblock \emph{Phys. Rev. Lett.}, 125:\penalty0 090503, Aug 2020.
\newblock \doi{10.1103/PhysRevLett.125.090503}.
\newblock URL \url{https://link.aps.org/doi/10.1103/PhysRevLett.125.090503}.

\bibitem[Hu et~al.(2016)Hu, Chen, Liu, Guo, Huang, Zhou, Han, Li, and
  Guo]{art:hu2016contextpathst}
Xiao-Min Hu, Jiang-Shan Chen, Bi-Heng Liu, Yu~Guo, Yun-Feng Huang, Zong-Quan
  Zhou, Yong-Jian Han, Chuan-Feng Li, and Guang-Can Guo.
\newblock Experimental test of compatibility-loophole-free contextuality with
  spatially separated entangled qutrits.
\newblock \emph{Physical Review Letters}, 117\penalty0 (17):\penalty0 170403,
  2016.

\bibitem[Hu et~al.(2018)Hu, Liu, Guo, Xiang, Huang, Li, Guo, Kleinmann,
  V\'ertesi, and Cabello]{art:CorrelationPath2018}
Xiao-Min Hu, Bi-Heng Liu, Yu~Guo, Guo-Yong Xiang, Yun-Feng Huang, Chuan-Feng
  Li, Guang-Can Guo, Matthias Kleinmann, Tam\'as V\'ertesi, and Ad\'an Cabello.
\newblock Observation of stronger-than-binary correlations with entangled
  photonic qutrits.
\newblock \emph{Phys. Rev. Lett.}, 120:\penalty0 180402, May 2018.
\newblock \doi{10.1103/PhysRevLett.120.180402}.
\newblock URL \url{https://link.aps.org/doi/10.1103/PhysRevLett.120.180402}.

\bibitem[Borges et~al.(2018)Borges, Baldij\~ao, Cond\'e, Cabral, Marques,
  Terra~Cunha, Cabello, and P\'adua]{art:expAutomatedOper}
G.~F. Borges, R.~D. Baldij\~ao, J.~G.~L. Cond\'e, J.~S. Cabral, B.~Marques,
  M.~Terra~Cunha, A.~Cabello, and S.~P\'adua.
\newblock Automated quantum operations in photonic qutrits.
\newblock \emph{Phys. Rev. A}, 97:\penalty0 022301, Feb 2018.
\newblock \doi{10.1103/PhysRevA.97.022301}.
\newblock URL \url{https://link.aps.org/doi/10.1103/PhysRevA.97.022301}.

\bibitem[Baldij\~ao et~al.(2017)Baldij\~ao, Borges, Marques,
  Sol\'{\i}s-Prosser, Neves, and P\'adua]{art:ThoreticalAutomatedOper}
R.~D. Baldij\~ao, G.~F. Borges, B.~Marques, M.~A. Sol\'{\i}s-Prosser, L.~Neves,
  and S.~P\'adua.
\newblock Proposal for automated transformations on single-photon multipath
  qudits.
\newblock \emph{Phys. Rev. A}, 96:\penalty0 032329, Sep 2017.
\newblock \doi{10.1103/PhysRevA.96.032329}.
\newblock URL \url{https://link.aps.org/doi/10.1103/PhysRevA.96.032329}.

\bibitem[Hecht(2002)]{liv:hecht2002optics}
E.~Hecht.
\newblock \emph{Optics}.
\newblock Addison-Wesley Longman, Incorporated, 2002.
\newblock ISBN 9780805385663.
\newblock URL \url{http://books.google.com.br/books?id=7aG6QgAACAAJ}.

\bibitem[Salazar-Serrano et~al.(2015)Salazar-Serrano, Valencia, and
  Torres]{art:torres}
Luis~José Salazar-Serrano, Alejandra Valencia, and Juan~P. Torres.
\newblock Tunable beam displacer.
\newblock \emph{Review of Scientific Instruments}, 86\penalty0 (3):\penalty0
  033109, 2015.
\newblock \doi{10.1063/1.4914834}.
\newblock URL \url{https://doi.org/10.1063/1.4914834}.

\bibitem[Teich and Saleh(1991)]{teich1991fundamentals}
Malvin~Carl Teich and BEA Saleh.
\newblock \emph{Fundamentals of photonics}, volume~3.
\newblock Wiley Interscience, 1991.

\bibitem[Monken et~al.(1998)Monken, Ribeiro, and P\'adua]{PhysRevA.57.3123}
C.~H. Monken, P.~H.~Souto Ribeiro, and S.~P\'adua.
\newblock Transfer of angular spectrum and image formation in spontaneous
  parametric down-conversion.
\newblock \emph{Phys. Rev. A}, 57:\penalty0 3123--3126, Apr 1998.
\newblock \doi{10.1103/PhysRevA.57.3123}.
\newblock URL \url{http://link.aps.org/doi/10.1103/PhysRevA.57.3123}.

\bibitem[Abouraddy et~al.(2001{\natexlab{a}})Abouraddy, Saleh, Sergienko, and
  Teich]{art:Saleh2001}
Ayman~F. Abouraddy, Bahaa E.~A. Saleh, Alexander~V. Sergienko, and Malvin~C.
  Teich.
\newblock Role of entanglement in two-photon imaging.
\newblock \emph{Phys. Rev. Lett.}, 87:\penalty0 123602, Aug 2001{\natexlab{a}}.
\newblock \doi{10.1103/PhysRevLett.87.123602}.
\newblock URL \url{https://link.aps.org/doi/10.1103/PhysRevLett.87.123602}.

\bibitem[Fonseca et~al.(1999)Fonseca, Ribeiro, P\'adua, and
  Monken]{PhysRevA.60.1530}
E.~J.~S. Fonseca, P.~H.~Souto Ribeiro, S.~P\'adua, and C.~H. Monken.
\newblock Quantum interference by a nonlocal double slit.
\newblock \emph{Phys. Rev. A}, 60:\penalty0 1530--1533, Aug 1999.
\newblock \doi{10.1103/PhysRevA.60.1530}.
\newblock URL \url{http://link.aps.org/doi/10.1103/PhysRevA.60.1530}.

\bibitem[de~Assis et~al.(2011)de~Assis, Carvalho, Berruezo, Ferraz, Santos,
  Sciarrino, and P\'{a}dua]{Pierre1}
P.-L. de~Assis, M.~A.~D. Carvalho, L.~P. Berruezo, J.~Ferraz, I.~F. Santos,
  F.~Sciarrino, and S.~P\'{a}dua.
\newblock Control of quantum transversecorrelations on a four-photon system.
\newblock \emph{Opt. Express}, 19\penalty0 (4):\penalty0 3715--3729, Feb 2011.
\newblock \doi{10.1364/OE.19.003715}.
\newblock URL \url{http://www.opticsexpress.org/abstract.cfm?URI=oe-19-4-3715}.

\bibitem[Sinha et~al.(2010)Sinha, Couteau, Jennewein, Laflamme, and
  Weihs]{art:Sinha418}
Urbasi Sinha, Christophe Couteau, Thomas Jennewein, Raymond Laflamme, and
  Gregor Weihs.
\newblock Ruling out multi-order interference in quantum mechanics.
\newblock \emph{Science}, 329\penalty0 (5990):\penalty0 418--421, 2010.
\newblock ISSN 0036-8075.
\newblock \doi{10.1126/science.1190545}.
\newblock URL \url{https://science.sciencemag.org/content/329/5990/418}.

\bibitem[Goodman(2005)]{liv:goodman2005}
J.W. Goodman.
\newblock \emph{Introduction to Fourier Optics}.
\newblock McGraw-Hill physical and quantum electronics series. Roberts \&
  Company Publishers, 2005.
\newblock ISBN 9780974707723.
\newblock URL \url{http://books.google.com.br/books?id=ow5xs\_Rtt9AC}.

\bibitem[Abouraddy et~al.(2001{\natexlab{b}})Abouraddy, Saleh, Sergienko, and
  Teich]{art:SchimdtConcurrence}
Ayman~F. Abouraddy, Bahaa E.~A. Saleh, Alexander~V. Sergienko, and Malvin~C.
  Teich.
\newblock Degree of entanglement for two qubits.
\newblock \emph{Phys. Rev. A}, 64:\penalty0 050101, Oct 2001{\natexlab{b}}.
\newblock \doi{10.1103/PhysRevA.64.050101}.
\newblock URL \url{https://link.aps.org/doi/10.1103/PhysRevA.64.050101}.

\bibitem[Neves et~al.(2007)Neves, Lima, Fonseca, Davidovich, and
  P\'adua]{LeoNevesetalConcurrence}
Leonardo Neves, G.~Lima, E.~J.~S. Fonseca, L.~Davidovich, and S.~P\'adua.
\newblock Characterizing entanglement in qubits created with spatially
  correlated twin photons.
\newblock \emph{Phys. Rev. A}, 76:\penalty0 032314, Sep 2007.
\newblock \doi{10.1103/PhysRevA.76.032314}.
\newblock URL \url{https://link.aps.org/doi/10.1103/PhysRevA.76.032314}.

\bibitem[Wootters(1998)]{WootersConcurrence}
William~K. Wootters.
\newblock Entanglement of formation of an arbitrary state of two qubits.
\newblock \emph{Phys. Rev. Lett.}, 80:\penalty0 2245--2248, Mar 1998.
\newblock \doi{10.1103/PhysRevLett.80.2245}.
\newblock URL \url{https://link.aps.org/doi/10.1103/PhysRevLett.80.2245}.

\bibitem[Cereceda(2003)]{cereceda2003degree}
Jos{\'e}~L Cereceda.
\newblock Degree of entanglement for two qutrits in a pure state.
\newblock \emph{arXiv preprint quant-ph/0305043}, 2003.

\bibitem[Rungta et~al.(2001)Rungta, Bu\ifmmode~\check{z}\else \v{z}\fi{}ek,
  Caves, Hillery, and Milburn]{art:Iconcurrenceintro}
Pranaw Rungta, V.~Bu\ifmmode~\check{z}\else \v{z}\fi{}ek, Carlton~M. Caves,
  M.~Hillery, and G.~J. Milburn.
\newblock Universal state inversion and concurrence in arbitrary dimensions.
\newblock \emph{Phys. Rev. A}, 64:\penalty0 042315, Sep 2001.
\newblock \doi{10.1103/PhysRevA.64.042315}.
\newblock URL \url{https://link.aps.org/doi/10.1103/PhysRevA.64.042315}.

\bibitem[Peres(2006)]{peres2006quantum}
Asher Peres.
\newblock \emph{Quantum theory: concepts and methods}, volume~57.
\newblock Springer Science \& Business Media, 2006.

\bibitem[Ekert and Knight(1995)]{ekert1995entangled}
Artur Ekert and Peter~L Knight.
\newblock Entangled quantum systems and the schmidt decomposition.
\newblock \emph{American Journal of Physics}, 63\penalty0 (5):\penalty0
  415--423, 1995.

\end{thebibliography}

\end{document}